\documentclass[reprint,aps,prc,superscriptaddress,nofootinbib]{revtex4}

\usepackage{amsmath,amssymb}
\usepackage{graphicx}
\usepackage{xcolor}
\usepackage{hyperref}
\usepackage{braket}
\usepackage{slashed}

\usepackage{comment}
\usepackage{bm}
\usepackage{fancyhdr}
\usepackage{float}
\usepackage{tikz}
\usepackage{multirow}
\usepackage{booktabs}
\usepackage{url}
\usepackage{siunitx}

\newcommand{\km}{K^{\!^-}\!\!({ \bar{u}} { s})}
\newcommand{\ph}{\phi({ s} { \bar{s}})}
\newcommand{\al}{\bar{\Lambda}(\bar{u}\bar{d}\bar{s})}
\newcommand{\pbar}{\bar{p}({ \bar{u}\bar{u}}{ \bar{d}})}
\newcommand{\ks}{\overline{\Xi}^{+}(\bar{d}\bar{s}\bar{s})}
\newcommand{\om}{{\Omega}^{-}(sss)}
\newcommand{\op}{\overline{\Omega}^{+}(\bar{s}\bar{s}\bar{s})}

\newcommand{\ols}[1]{\mskip.5\thinmuskip\overline{\mskip-.5\thinmuskip {#1} \mskip-.5\thinmuskip}\mskip.5\thinmuskip} 

\begin{document}

\title{
Electric-charge-dependent directed flow splitting of produced quarks in Au+Au collisions}


\affiliation{Academia Sinica}
\affiliation{Abilene Christian University, Abilene, Texas   79699}
\affiliation{AGH University of Krakow, FPACS, Cracow 30-059, Poland}
\affiliation{Alikhanov Institute for Theoretical and Experimental Physics NRC "Kurchatov Institute", Moscow 117218}
\affiliation{Argonne National Laboratory, Argonne, Illinois 60439}
\affiliation{American University in Cairo, New Cairo 11835, Egypt}
\affiliation{Ball State University, Muncie, Indiana, 47306}
\affiliation{Brookhaven National Laboratory, Upton, New York 11973}
\affiliation{University of Calabria \& INFN-Cosenza, Rende 87036, Italy}
\affiliation{University of California, Berkeley, California 94720}
\affiliation{University of California, Davis, California 95616}
\affiliation{University of California, Los Angeles, California 90095}
\affiliation{University of California, Riverside, California 92521}
\affiliation{Central China Normal University, Wuhan, Hubei 430079 }
\affiliation{University of Illinois at Chicago, Chicago, Illinois 60607}
\affiliation{Chongqing University, Chongqing, 401331}
\affiliation{Creighton University, Omaha, Nebraska 68178}
\affiliation{Czech Technical University in Prague, FNSPE, Prague 115 19, Czech Republic}
\affiliation{National Institute of Technology Durgapur, Durgapur - 713209, India}
\affiliation{ELTE E\"otv\"os Lor\'and University, Budapest, Hungary H-1117}
\affiliation{Frankfurt Institute for Advanced Studies FIAS, Frankfurt 60438, Germany}
\affiliation{Fudan University, Shanghai, 200433 }
\affiliation{Guangxi Normal University, Guilin, 541004}
\affiliation{University of Heidelberg, Heidelberg 69120, Germany }
\affiliation{University of Houston, Houston, Texas 77204}
\affiliation{Huzhou University, Huzhou, Zhejiang  313000}
\affiliation{Indian Institute of Science Education and Research (IISER), Berhampur 760010 , India}
\affiliation{Indian Institute of Science Education and Research (IISER) Tirupati, Tirupati 517507, India}
\affiliation{Indian Institute Technology, Patna, Bihar 801106, India}
\affiliation{Indiana University, Bloomington, Indiana 47408}
\affiliation{Institute of Modern Physics, Chinese Academy of Sciences, Lanzhou, Gansu 730000 }
\affiliation{University of Jammu, Jammu 180001, India}
\affiliation{Joint Institute for Nuclear Research, Dubna 141 980}
\affiliation{Kent State University, Kent, Ohio 44242}
\affiliation{University of Kentucky, Lexington, Kentucky 40506-0055}
\affiliation{Lanzhou University}
\affiliation{Lawrence Berkeley National Laboratory, Berkeley, California 94720}
\affiliation{Lehigh University, Bethlehem, Pennsylvania 18015}
\affiliation{Max-Planck-Institut f\"ur Physik, Munich 80805, Germany}
\affiliation{Michigan State University, East Lansing, Michigan 48824}
\affiliation{National Research Nuclear University MEPhI, Moscow 115409}
\affiliation{National Institute of Science Education and Research, HBNI, Jatni 752050, India}
\affiliation{National Cheng Kung University, Tainan 70101 }
\affiliation{The Ohio State University, Columbus, Ohio 43210}
\affiliation{Panjab University, Chandigarh 160014, India}
\affiliation{NRC "Kurchatov Institute", Institute of High Energy Physics, Protvino 142281}
\affiliation{Purdue University, West Lafayette, Indiana 47907}
\affiliation{Rice University, Houston, Texas 77251}
\affiliation{Rutgers University, Piscataway, New Jersey 08854}
\affiliation{University of Science and Technology of China, Hefei, Anhui 230026}
\affiliation{South China Normal University, Guangzhou, Guangdong 510631}
\affiliation{Sejong University, Seoul, 05006, South Korea}
\affiliation{Shandong University, Qingdao, Shandong 266237}
\affiliation{Shanghai Institute of Applied Physics, Chinese Academy of Sciences, Shanghai 201800}
\affiliation{Southern Connecticut State University, New Haven, Connecticut 06515}
\affiliation{State University of New York, Stony Brook, New York 11794}
\affiliation{Instituto de Alta Investigaci\'on, Universidad de Tarapac\'a, Arica 1000000, Chile}
\affiliation{Temple University, Philadelphia, Pennsylvania 19122}
\affiliation{Texas A\&M University, College Station, Texas 77843}
\affiliation{University of Texas, Austin, Texas 78712}
\affiliation{Tsinghua University, Beijing 100084}
\affiliation{University of Tsukuba, Tsukuba, Ibaraki 305-8571, Japan}
\affiliation{University of Chinese Academy of Sciences, Beijing, 101408}
\affiliation{Valparaiso University, Valparaiso, Indiana 46383}
\affiliation{Variable Energy Cyclotron Centre, Kolkata 700064, India}
\affiliation{Wayne State University, Detroit, Michigan 48201}
\affiliation{Wuhan University of Science and Technology, Wuhan, Hubei 430065}
\affiliation{Yale University, New Haven, Connecticut 06520}

\author{B.~E.~Aboona}\affiliation{Texas A\&M University, College Station, Texas 77843}
\author{J.~Adam}\affiliation{Czech Technical University in Prague, FNSPE, Prague 115 19, Czech Republic}
\author{G.~Agakishiev}\affiliation{Joint Institute for Nuclear Research, Dubna 141 980}
\author{I.~Aggarwal}\affiliation{Panjab University, Chandigarh 160014, India}
\author{M.~M.~Aggarwal}\affiliation{Panjab University, Chandigarh 160014, India}
\author{Z.~Ahammed}\affiliation{Variable Energy Cyclotron Centre, Kolkata 700064, India}
\author{A.~Aitbaev}\affiliation{Joint Institute for Nuclear Research, Dubna 141 980}
\author{I.~Alekseev}\affiliation{Alikhanov Institute for Theoretical and Experimental Physics NRC "Kurchatov Institute", Moscow 117218}\affiliation{National Research Nuclear University MEPhI, Moscow 115409}
\author{E.~Alpatov}\affiliation{National Research Nuclear University MEPhI, Moscow 115409}
\author{A.~Aparin}\affiliation{Joint Institute for Nuclear Research, Dubna 141 980}
% ...

\collaboration{STAR Collaboration}\noaffiliation

\date{\today}

\begin{abstract}

We report directed flow ($v_1$) of multistrange baryons ($\Xi$ and $\Omega$) and improved $v_1$ data for $K^{-}$, $\bar{p}$, $\bar{\Lambda}$ and $\phi$ in Au+Au collisions at $\sqrt{s_{\mathrm{NN}}}=$27 and 200 GeV from the STAR experiment at the Relativistic Heavy Ion Collider (RHIC). We focus on particles whose constituent quarks are not transported from the incoming nuclei but instead are produced in the collisions. At intermediate impact parameters, we examine quark coalescence behavior for particle combinations with identical quark content, and search for any departure from this behavior (``splitting'') for combinations having non-identical quark content. Under the assumption of quark coalescence for produced quarks, the splitting strength appears to increase with the electric charge difference of the constituent quarks in the combinations, consistent with electromagnetic effect expectations. 

%

\end{abstract}

\pacs{}

\keywords{Directed flow, Electric charge, Strangeness, Heavy-ion collisions, Electromagnetic field} 


\maketitle

The first harmonic in the Fourier expansion of the azimuthal distribution of emitted particles relative to the reaction plane in a nucleus-nucleus collision~\cite{Voloshin:1994mz,Poskanzer:1998yz,Bilandzic:2010jr} is known as directed flow, $v_1$, a collective sideward motion whose dominant component is an odd function of the particle rapidity ($y$). The  sign convention of $v_1(y)$ is that the projectile and target fragments/spectators have positive $v_1$ at positive rapidity ($y>0$), and negative $v_1$ at negative rapidity ($y<0$).
The early collision dynamics~\cite{Kolb:2003dz,Huovinen:2006jp,Sorge:1996pc} can be probed by $v_1(y)$, an interpretation supported by hydrodynamic~\cite{Heinz:2009xj} and nuclear transport~\cite{Bass:1998ca} models.

The early stage of these collisions features a magnetic field on the order of $10^{14}-10^{15}$ T~\cite{Skokov:2009qp}, dominated by the passing spectator protons. While this magnetic field strength well exceeds that produced in any other context, its time evolution and role during a heavy-ion collision greatly depends on the electric conductivity of the medium. Experimentally,  efforts to date to constrain the electric conductivity, and consequently to probe the electromagnetic field, have fallen short. In this Letter, we investigate the use of directed flow splitting measurements to probe possible signatures of the early-stage electromagnetic field in Au+Au collisions.

 The term splitting is typically used in the literature to refer to a difference in flow between a particle and its antiparticle \cite{Xu:2013sta,Xu:2016ihu,Voronyuk:2014rna,Toneev:2016bri,Oliva:2019kin,Das:2016cwd, Chatterjee:2017ahy, Chatterjee:2018lsx,Adamczyk:2016eux,Acharya:2019ijj,Adam:2019wnk}. In heavy-ion collisions, EM-field effects manifest as a combination of Faraday, Coulomb and Hall effect~\cite{Gursoy:2018yai,Gursoy:2014aka,Dubla:2020bdz}. The combined effect of EM fields drives positive and negative charges in opposite directions, resulting in the splitting of their $v_1$. The present analysis studies splitting in the more general case of a difference in $v_1$ between particle combinations with different charge~\cite{Sheikh:2021rew}. 

 
As the spectator protons recede from the collision zone, the magnetic field decreases quickly and the resulting Faraday induction produces an electric current. The charged spectators also exert a Coulomb force on the charged constituents. Meanwhile, the collision zone expands in the longitudinal (beam) direction, which is on average perpendicular to the magnetic field. The Lorentz force acts perpendicular to both the longitudinal velocity and the magnetic field, analogous to the Hall effect~\cite{Gursoy:2018yai,Gursoy:2014aka}. The combination of Faraday, Coulomb and Hall effect may influence $v_1$ of the emitted particles~\cite{Gursoy:2018yai,Gursoy:2014aka,Dubla:2020bdz}.

\begin{figure}[htb]
    \centering
   \includegraphics[width=0.47\textwidth]{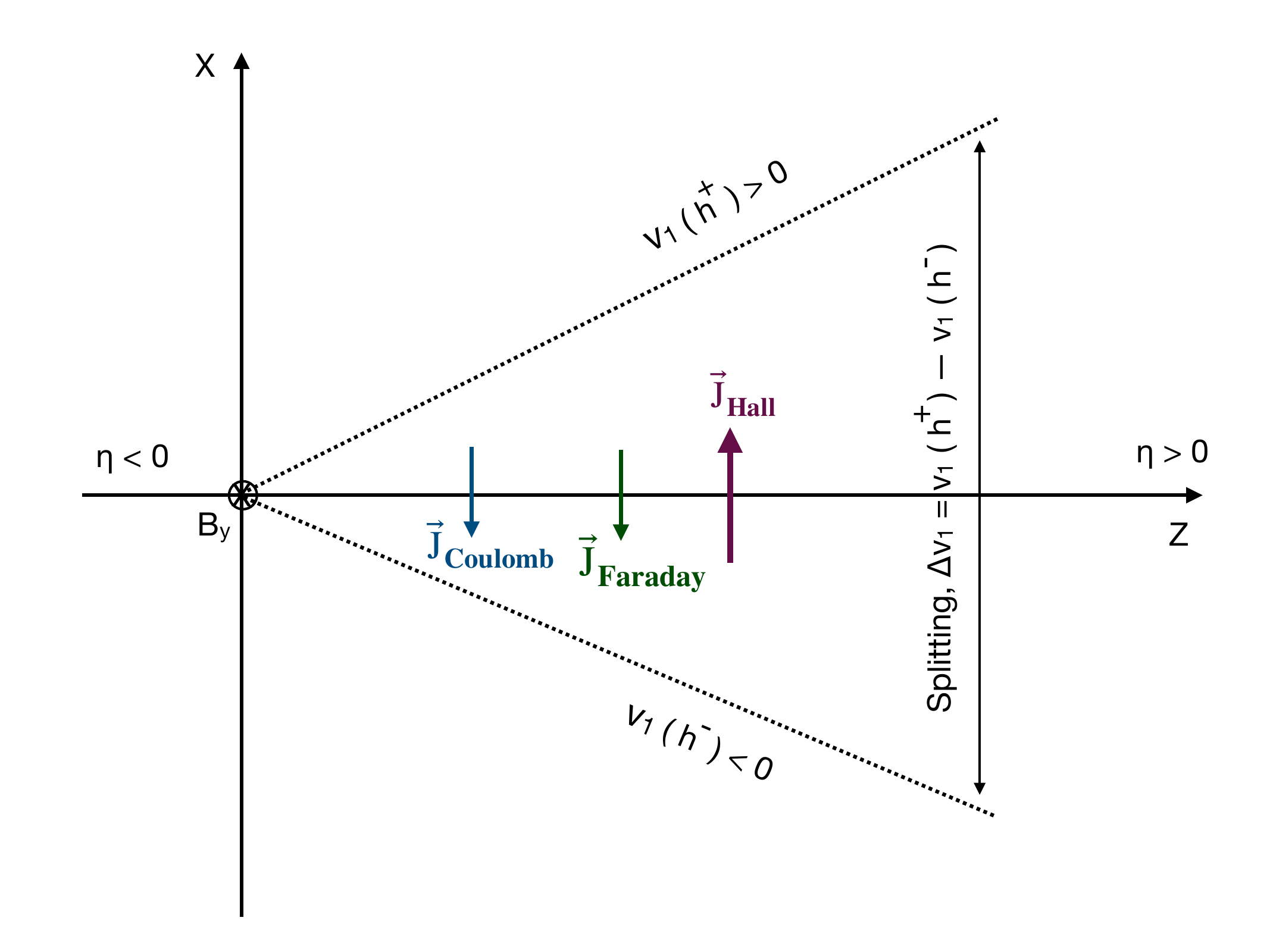}
    \caption{Diagram (motivated by Refs.~\cite{Gursoy:2018yai,Gursoy:2014aka}) illustrating the influence of the magnetic field $\vec{B}$ on $v_1$ of electric charges. 
    The colliding nuclei at $+z$ ($-z$) are located at $+x$ ($-x$), and the generated $\vec{B}$ is along $-y$.}
    \label{cartoon}
 \end{figure}

The sign of $v_1$ splitting can reveal which aspect of the EM field is dominant. When the Hall effect overcomes the Faraday and Coulomb effects, the $v_1$ of positive charges ($h^+$) at $y>0$ becomes positive, and the opposite happens for $h^-$. 
This causes the splitting in $v_1$, as illustrated in Fig.~\ref{cartoon}. The produced medium expands longitudinally. The directions of the resultant electric currents due to Faraday, Hall and Coulomb effects are shown. The $v_1$ directions for $h^{+}$ and $h^{-}$ are illustrated for the case where the Hall current exceeds the Faraday+Coulomb current.  
When the Faraday+Coulomb effect is stronger, the $v_1$ directions for $h^+$ and $h^-$ are reversed. The $v_1$ splitting between opposite-charge light hadron pairs like $\pi^\pm$, $K^\pm$, $p-\bar{p}$, etc.~\cite{Gursoy:2018yai,Voronyuk:2014rna,Toneev:2016bri,Oliva:2019kin} as well as heavy pairs like $D^{0}\,(c\bar{u})$ and $\overline{D^{0}}\,(\bar{c}u)$~\cite{Das:2016cwd,Chatterjee:2017ahy,Chatterjee:2018lsx} has been predicted. The predicted splitting 
for the latter is stronger, 
suggesting dominance of the magnetic field on heavy quarks due to their early production and large relaxation time in the medium~\cite{Das:2016cwd,Chatterjee:2017ahy,Chatterjee:2018lsx}. 

Charge-dependent $v_1$ splitting between $h^{+}$ and $h^{-}$ in Cu+Au and Au+Au collisions at $\sqrt{s_{\mathrm{NN}}}=200$ GeV was reported by STAR~\cite{Adamczyk:2016eux}. A large $v_1$ splitting was observed in Cu+Au, attributed to the stronger Coulomb force in mass-asymmetric Cu+Au compared to Au+Au. A $v_1$ splitting with a
significance of 2.6$\sigma$ in Pb+Pb at $\sqrt{s_{\mathrm{NN}}}=5.02$ TeV was observed by the ALICE Collaboration~\cite{Acharya:2019ijj}. Moreover, a large $v_1$ for $D^0$, $\ols{D}^{0}$ mesons and a nonzero $v_1$ splitting between $D^0$ and $\ols{D}^{0}$ was reported by STAR~\cite{Adam:2019wnk} in $\sqrt{s_{\mathrm{NN}}}=200$ GeV Au+Au and by ALICE~\cite{Acharya:2019ijj} in $\sqrt{s_{\mathrm{NN}}}=5.02$ TeV Pb+Pb. During STAR Beam Energy Scan~\cite{STAR:2010vob} data collection, the Heavy Flavor Tracker~\cite{Contin:2017mck} was absent, and the small heavy flavor production rate at those beam energies makes it unfeasible to measure splitting for $D^0/\ols{D}^{0}$, hence the present focus is on strange particles. 

Production of $s$ and $\bar{s}$ quarks is enhanced when nuclei collide with sufficient energy to produce a quark-gluon plasma \cite{Rafelski:1982pu}, yielding relatively abundant strange particles like $K$, $\Lambda$, $\phi$, $\Xi$ and $\Omega$~\cite{Rafelski:1982pu,Shor:1984ui}.  Due to their low scattering cross sections, thermal freezeout occurs earlier for the multistrange baryons $\Xi$ and $\Omega$ than for particles containing one or zero strange quarks~\cite{vanHecke:1998yu}; being emitted near the phase boundary of the hadronising fireball, they carry important information about the early stages of the collisions~\cite{Adams:2003fy}. 

The collision fireball can have a large net-baryon number, depending on the collision energy. There could be increased splitting with strangeness and baryon content, suggesting a potential role for the initial baryon inhomogeneity 
\cite{Parida:2023ldu,Bozek:2022svy,Parida:2022zse,Parida:2022ppj} in the observed splitting, and deserves further investigation. 

Hadrons containing $u$ and $d$ quarks can be either transported from the initial nuclei~\cite{Dunlop:2011cf} or produced in the collisions. Transported quarks
undergo more interactions, and have different $v_1$~\cite{Guo:2012qi} from the produced $u$ and $d$ quarks or other produced quarks ($\bar{u}$, $\bar{d}$, $s$ and $\bar{s}$) that do not have a pre-collision history. Produced quarks are generated through partonic fusion during collisions. This dependence on origin among $u$ and $d$ quarks complicates the interpretation of splitting arising from EM fields~\cite{Sheikh:2021rew}, and are therefore excluded from the present analysis. STAR's recent observation of proton $v_1$ splitting~\cite{STAR:2023jdd}, exhibiting a sign change from central to peripheral collisions, suggests a predominance of Faraday+Coulomb effects over Hall effect in peripheral events, provided that the $v_1$ splitting of transported quarks is independent of centrality.

Various models suggest that collective flow mostly develops before hadronisation, with particles forming via coalescence of constituent quarks of near-equal velocity.
In this scenario, in the limit of small azimuthal anisotropy $v_n$, the $v_n$ of the resulting hadrons equal the summed $v_n$ for their constituent quarks: $v_1({\rm hadron}) = \sum\limits_i v_1(q_i)$. This is known as the coalescence sum rule, and the number-of-constituent-quark (NCQ) scaling, an approximate scaling
that comes from the addition of the valence-quark momenta at hadronisation, follows from it \cite{Dunlop:2011cf}. However,  several theoretical considerations suggest that the NCQ scaling should be violated in certain
conditions. For instance, the inclusion of higher Fock states, which account for the contribution of sea quarks and gluons, has been shown to influence NCQ scaling~\cite{Muller:2005pv}. Similarly, models that incorporate the recombination of "thermal" partons (soft partons thermalized within the medium) and "shower" partons (partons originating from jet fragmentation) predict centrality-dependent deviations from NCQ scaling that vary by particle species~\cite{Chiu:2008lzv}. There are experimental measurements at RHIC~\cite{PHENIX:2012swz} and LHC~\cite{ALICE:2014wao} showing the deviations of NCQ scaling for elliptic flow, $v_2$. If the collective flow has a substantial hadronic contribution originating after hadronisation, the simple sum rule might no longer apply. The present work utilizes $v_1$, which is widely accepted to develop earlier than $v_2$. The sum rule for $v_1$ is observed to hold to a better level of accuracy if transported quarks are avoided, as in the current study.
Any experimental test of the coalescence sum rule is problematic when $u$ and $d$ quarks are involved, because of the difficulty in distinguishing between transported and produced quarks~\cite{Sheikh:2021rew,Nayak:2023ofv}.  
Seven qualifying particle species have adequate experimental statistics: $\km$, $\pbar$, $\al$, $\ph$, $\ks$, $\om$ and $\op$. Since $v_1$ is likely sensitive to the constituent quark mass, we combine various particle species into a pair of groups having zero or close-to-zero total mass difference ($\Delta m$) at the constituent quark level. However, specific values are not assigned to quark masses. Instead, we ensure equivalent strange-quark content on each side of the minus sign in our combinations. The expression $\Delta m =0$ signifies the balancing of strange quarks, without requiring any assumption about the mass of individual quarks.

We then study the splitting of $v_1$ as a function of charge difference between group pairs ($\Delta q$), which should be sensitive to EM effects, as well as strangeness difference ($\Delta S$). Of course, $\Delta S$ is not independent of $\Delta q$. With these seven particle species, there are five independent combinations \cite{Sheikh:2022gbf}. The identical quark combination case is $\Delta v_{1} (\Delta q=0,\, \Delta S=0) = \{v_1[\pbar] + v_1[\ph]\} - \{v_1[\km] + v_1[\al]\}$. The other four independent cases involve nonidentical quark combinations and are listed in Table~\ref{tab:delq_dels} (Appendix~\ref{a1}).  

This Letter outlines the measurements of $v_1(y)$ for $\Xi$ and $\Omega$ in Au+Au collisions at $\sqrt{s_{\mathrm{NN}}} = 27$ and $200$ GeV, now a feasible measurement by STAR at RHIC. We analyse minimum-bias trigger data for Au+Au collisions at $\sqrt{s_{\mathrm{NN}}}=$ 27 and 200 GeV. Results for 10-40\% centrality are reported, where $v_1$ is the strongest, while 40-80\% centrality is covered in the supplemental material. We investigate the $v_1$ splitting between various quark and antiquark combinations versus $\Delta q$ and $\Delta S$.  


The Event Plane Detectors (EPD) covering pseudorapidity $2.1 < |\eta| < 5.1$~\cite{Adams:2019fpo} and the Zero-Degree Calorimeter Shower-Maximum Detectors (ZDC-SMD) covering $|\eta|>6.3$~\cite{Adamczyk:2011aa} determine the event plane at $\sqrt{s_{\mathrm{NN}}}=$ 27 and 200 GeV, respectively. 
The STAR Time Projection Chamber (TPC)~\cite{Anderson:2003ur} is used for charged particle tracking within $|\eta|<1$ and $p_\mathrm{T}\!>\!0.2$ GeV/$c$. Typical  tracking efficiencies (estimated using Geant-based embedding) for protons and antiprotons vary between 60–80\% and 80–83\% at the most central (0–5\%) and peripheral (70–80\%) centralities, respectively, at low $p_T$ (0.4 < $p_T$ < 0.8 GeV/$c$)~\cite{STAR:2019ans}. For centrality determination, the probability distributions of uncorrected TPC tracks within $|\eta| < $ 0.5 are used. The centrality classes are based on fits to Monte-Carlo Glauber simulations~\cite{STAR:2008med, Miller:2007ri, STAR:2021mii}. The longitudinal position ($V_{z}$) and the radial position ($V_r$) of the primary vertex are reconstructed from TPC tracks. We require $V_{r}<2$ cm, while $|V_{z}|<70$ cm at 27 GeV and $|V_{z}|<30$ cm at 200 GeV. Tracks must satisfy $p_\mathrm{T}\!>\!0.2$ GeV/$c$ and a distance of closest approach from the primary vertex under 3 cm. We also require tracks to have at least 15 space points in the TPC acceptance ($|\eta|<1$) and the number of measured space points must be greater than 0.52 times the maximum possible number. Particle identification is based on energy loss in the TPC and time from the Time Of Flight detector~\cite{Bonner:2003bv,Llope:2012zz}. Reliable identification requires $0.4\!<\!p_\mathrm{T}\!<\!5$ GeV/$c$ for $p$ and $\bar{p}$, and total momentum $p$ $\!<\!1.6$ GeV/$c$ for $\pi^\pm$ and $K^\pm$. 

The $\Lambda$, $\bar{\Lambda}$, ${\Xi}^{-}$, $\overline{\Xi}^{+}$, ${\Omega}^{-}$ and $\overline{\Omega}^{+}$ candidates within $\!0.2<p_\mathrm{T}\!<\!5$ GeV/$c$ are reconstructed by the Kalman filter (KF) method 
\cite{Gorbunov:2013phd,Zyzak:2016phd,Kisel:2018nvd,ALICE:2021bli,ALICE:2022exq,ALICE:2022cop}. 
The KF exploits the track fit quality and decay topology. We reconstruct $\Lambda$ ($\bar{\Lambda}$), $\Xi^{-}$ ($\overline{\Xi}^{+}$) and ${\Omega}^{-}$ ($\overline{\Omega}^{+}$) via $p\pi^{-}$ ($\bar{p}\pi^{+}$), $\Lambda\pi^{-}$ ($\bar{\Lambda}\pi^{+}$) and $\Lambda K^-$ ($\bar{\Lambda}K^+$) decay channels, respectively, with $>$96\% purity~\cite{STAR:2020xbm}. The statistical significance of short-lived particle reconstruction by the KF is improved by $\sim$30\% for $\Xi^{-}$ ($\overline{\Xi}^{+}$) compared to the standard 
method~\cite{STAR:2020xbm,Adam:2018ivw,Adam:2019koz}. The KF also ensures that hyperon candidates do not share daughters with other particles of interest. The $\phi$ meson decay to $K^+K^-$ is a strong process that results in a large background/signal ratio and hence the KF cannot provide a pure $\phi$ sample. The $\phi$ mesons within $0.2<p_\mathrm{T}\!<\!10$ GeV/$c$ are reconstructed via the invariant mass technique with background subtraction by pair rotation~\cite{ALICE:2016yta}. The $v_1$ for $K^+K^-$ pairs (signal+background) is estimated as a function of pair invariant mass, $m_{\rm inv}$, in various $y$ bins, $v_1^{\rm Sig+Bkg} (m_{\rm inv})$, which is then decomposed into two parts:
$v_1^{\rm Sig+Bkg} (m_{\rm inv}) = Y_R v_1^{\rm Sig} (m_{\rm inv}) + (1-Y_R) v_1^{\rm Bkg} (m_{\rm inv})$, where $Y_R = {\rm Yield(Sig)}/[{\rm Yield(Sig)+Yield(Bkg)}]$. The relative signal yields are extracted by fitting the invariant mass distribution, where $v_1^{\rm Bkg}$ is the background $v_1$ parametrized as a first-order polynomial, and $v_1^{\rm Sig}$ is the $v_1$ for $\phi$ mesons.

The $v_1$ is determined using the event plane method: $v_1 = \langle \cos(\phi-\Psi_1)\rangle/{\rm Res}\{\Psi_1\}$, where $\phi$ is the track azimuth, $\Psi_1$ is the first-order event plane and ${\rm Res}\{\Psi_1\}$ is its resolution~\cite{Poskanzer:1998yz,STAR:2005btp,STAR:2011hyh}. For 10-40\% centrality, ${\rm Res}\{\Psi_1\}$ is 0.494 and 0.366 at $\sqrt{s_{\mathrm{NN}}}=$ 27 and 200 GeV, respectively. We evaluate $v_1$ differences $\Delta v_1$ for all 5 combination pairs, keeping the same ($p_\mathrm{T}/n_q,~y$) region (where $n_q=2$ for mesons and 3 for baryons), namely $0.13 \!<\! (p_\mathrm{T}/n_q) \!<\! 1$ GeV/$c$ and $|y| \!<\! 0.8$. For each combination pair, the strength of splitting is characterized by the slope $F_{\Delta}=d\Delta v_1/dy$. $F_{\Delta}$ is extracted from a linear fit of the measured $\Delta v_1$ for all pair combinations. 

Systematic uncertainties are estimated by varying the $z$-vertex, track quality and PID-related selection criteria. We use Barlow's method~\cite{Barlow:2002yb} to remove statistical fluctuations from the systematic uncertainty. Typical systematic uncertainty in $v_1(y)$, $\Delta v_1(y)$ and $d\Delta v_1/dy$ due to the event, track, and topological selection criteria variations combined is below 5\%. We also account for systematic uncertainties arising from trigger bias and beam luminosity changes, estimated to be under 2\%. Corrections for TPC tracking efficiency and acceptance are small, and are included in the systematic uncertainty. Systematic uncertainties from all sources are added in quadrature. Systematic uncertainties on slopes are obtained by fitting for all possible systematic variations and using Barlow's method to obtain final systematic uncertainties. Nonflow (azimuthal correlations unrelated to the reaction plane orientation, arising from resonances, jets, quantum statistics, and final-state interactions) is minimized by the sizable pseudorapidity gap between the TPC and the EPD/ZDC-SMD~\cite{Ollitrault:1992bk,Voloshin:1994mz,Poskanzer:1998yz,Bilandzic:2010jr,Agakishiev:2011id,Adamczyk:2014ipa,Adamczyk:2017nxg}. We investigated the effect of the pseudorapidity gap on the $v_1$ measurements using the inner EPD, which has a pseudorapidity coverage of 3.4 < |$\eta$| < 5, resulting in a gap of 2.4 units with the TPC, compared to a smaller gap of 1.1 units with the full EPD. The $v_1$ measurements for antiprotons obtained with both the full and inner EPD overlap within the marker size, with $v_1$-slopes of $-$0.024 $\pm$ 0.001 and $-$0.022 $\pm$ 0.001, respectively.

\begin{figure}[htb]
    \centering
   \includegraphics[width=0.49\textwidth]{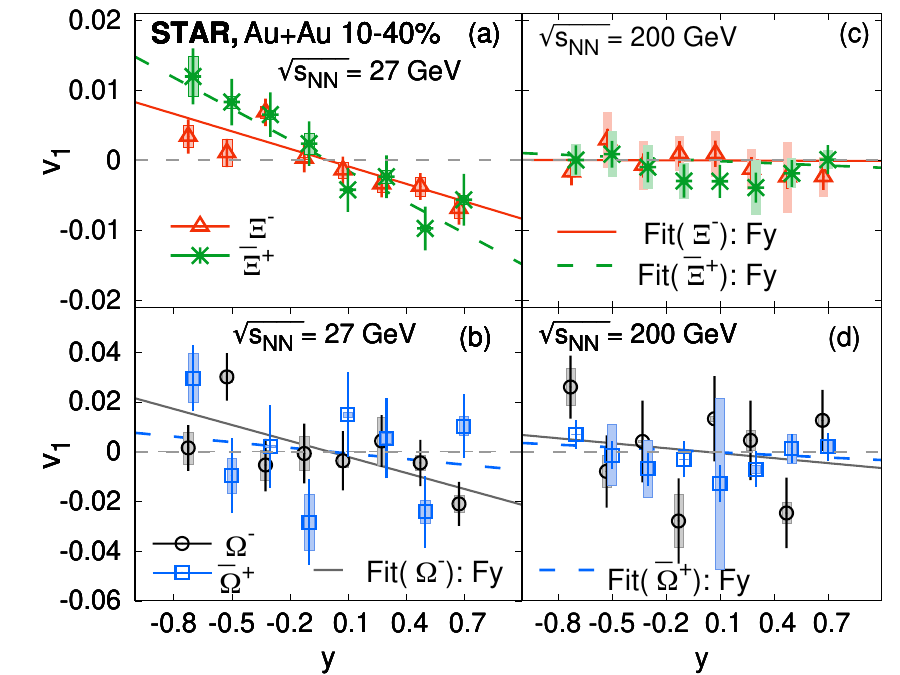}
   \caption{Directed flow of $\Xi^{-}$, $\overline{\Xi}^{+}$, ${\Omega}^{-}$ and $\overline{\Omega}^{+}$ versus rapidity for 10-40\% Au+Au at $\sqrt{s_{\mathrm{NN}}} = 27$ GeV (a, b) and 200 GeV (c, d). The bars and shaded bands denote statistical and systematic uncertainties, respectively. Points for ${\Omega}^{-}$ and $\Xi^{-}$ are staggered horizontally for better visualization.}
   \label{fig:v1_xiom}
\end{figure}

\begin{figure}
    \centering
   \includegraphics[width=0.49\textwidth]{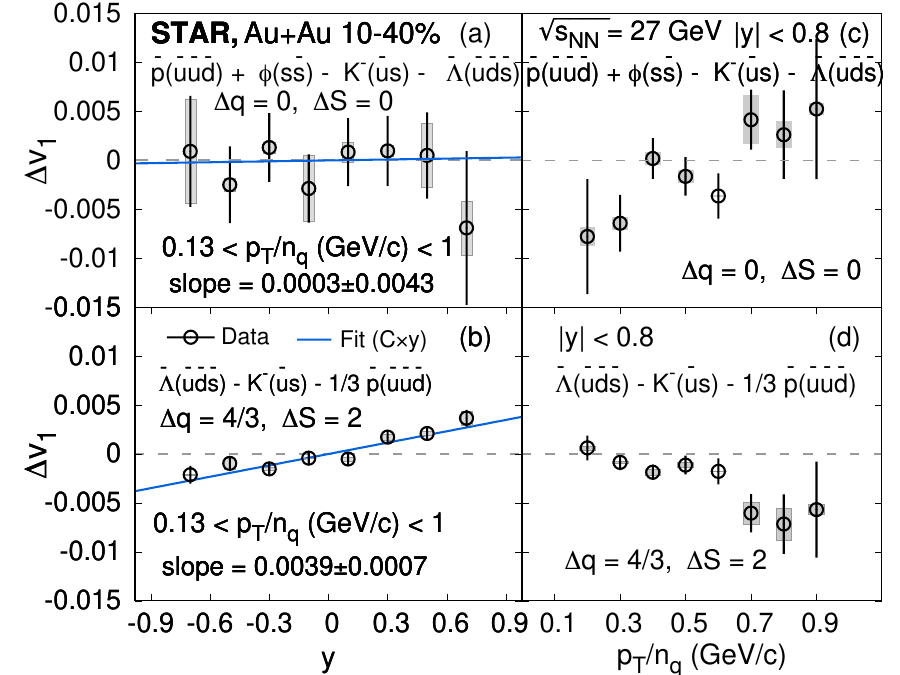}
   \caption{$\Delta v_1$ versus rapidity (a, b) and $p_\mathrm{T}/n_q$ (c, d) for ($\Delta q,\,\Delta S$)=(0,~0) and (4/3,~2) in 10-40\% Au+Au at $\sqrt{s_{\mathrm{NN}}}=27$ GeV. 
   }
   \label{fig:delv1y}
\end{figure}

\begin{figure}
   \centering
    \includegraphics[width=0.49\textwidth]{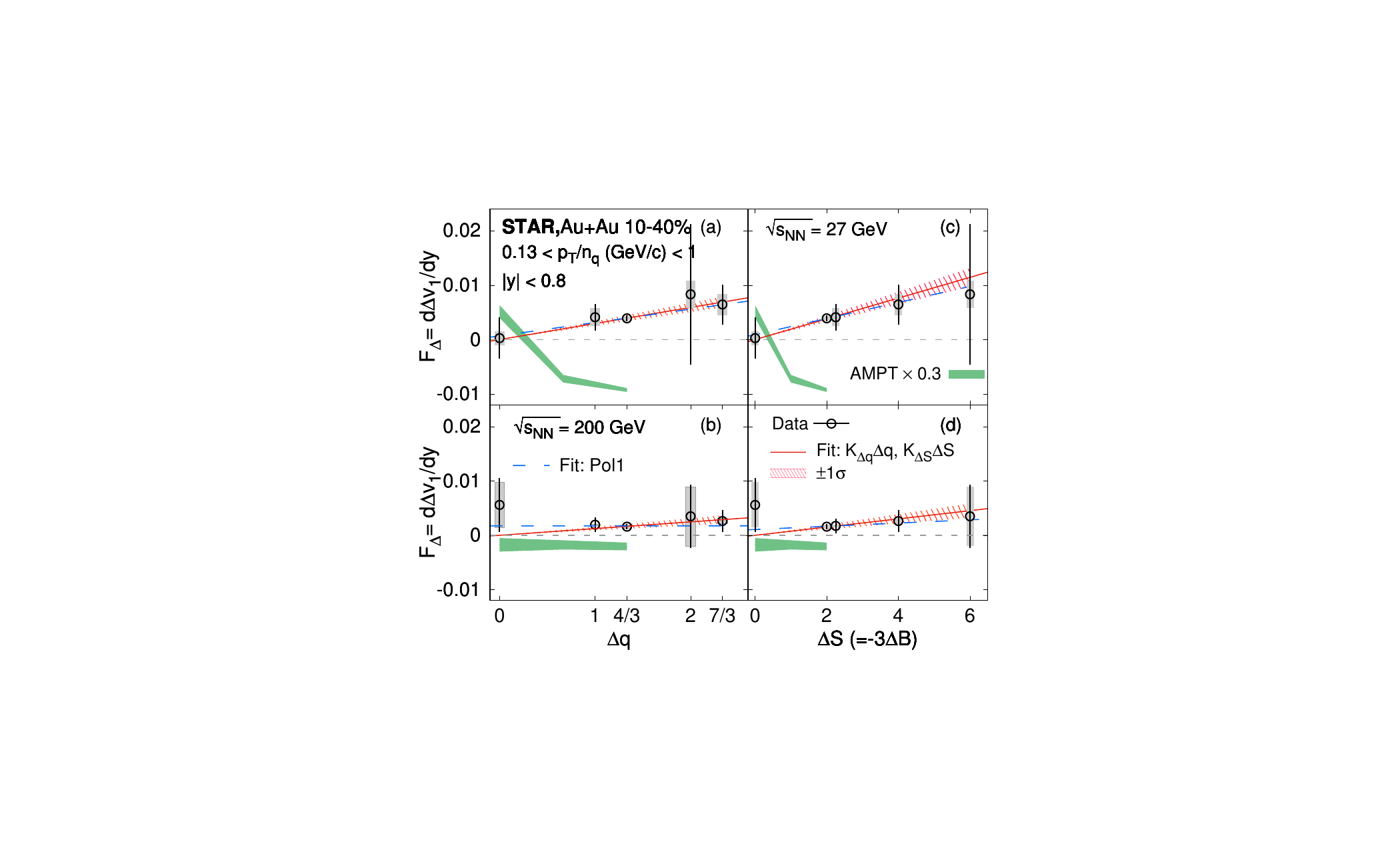}
    \caption{ Midrapidity $\Delta v_1$ slope versus $\Delta q$ (a, b) and $\Delta S$ (c, d) in 10-40\% Au+Au at 27 GeV (a, c) and 200 GeV (b, d). The red line is a linear fit through the origin, and the dashed blue line (labeled Pol1) is a linear fit allowing a non-zero intercept. Two points lie at $\Delta S=2$; one is displaced horizontally for clarity.  Solid green bands show AMPT from Ref.~\cite{Nayak:2019vtn}, 
  where the band's width denotes the model's statistical uncertainty. 
    }
  \label{fig:slope_delq_dels}
\end{figure}

Figure~\ref{fig:v1_xiom} reports $\Xi$ and ${\Omega}$ baryon $v_1$ for 10-40\% centrality in Au+Au at $\sqrt{s_{\mathrm{NN}}} = 27$ and 200 GeV.  A linear fit $v_1(y)=Fy$ yields slope parameters
$10^{4}\times F=-214\pm79 
 \mathrm{(stat.)}\pm34 \mathrm{(syst.)}$ $[-75\pm120 \mathrm{(stat.)}\pm17 \mathrm{(syst.)}]$ for $\Omega^{-}$ [$\overline{\Omega}^{+}$] and $10^{4}\times F=-83\pm20 \mathrm{(stat.)} \pm0 \mathrm{(syst.)}$ $[-150\pm28\mathrm{(stat.)}\pm13 \mathrm{(syst.)}]$ for $\Xi^{-}$ [$\overline{\Xi}^{+}$] at $\sqrt {s_{\mathrm{NN}}} = 27$ GeV. These quoted errors are absolute. 
 The $v_1$ of ${\Xi}^{-}$ is not used, because ${\Xi}^{-}$ contains the possibly-transported quark $d$, whereas $v_1$ of $\overline{\Xi}^{+}$, which has only produced quarks, is included in our combination index 5 (see Table~\ref{tab:delq_dels} of Appendix~\ref{a1}).

We first test the coalescence sum rule with identical quark combinations, $\it{i.e.}$, the equality $v_1[\pbar] + v_1[\ph] = v_1[\km] + v_1[\al]$ in a common kinematic region ($p_\mathrm{T}/n_q$,~$y$). On both sides, the constituent quark content is $\bar{u}\bar{u}\bar{d}s\bar{s}$, so $\Delta m=0$, $\Delta q=0$ and $\Delta S=0$. 
Then we measure $\Delta v_1$ for nonidentical quark combinations $\Delta m\approx 0$, $\Delta q\ne0$, and $\Delta S\ne0$. Figure~\ref{fig:delv1y} shows the measured $\Delta v_1$ for ($\Delta q,\,\Delta S$)= (0,~0) in panels (a) and (c) and for (4/3,~2) in panels (b) and (d) at 10-40\% centrality in Au+Au at $\sqrt{s_{\mathrm{NN}}} = 27$ GeV. 
We fit a slope $F_\Delta$ assuming $\Delta v_1(y)=F_\Delta y$. For ($\Delta q,\,\Delta S$) = (0,~0), 
the fitted slope is consistent with zero at 10-40\% centrality and is about $2\sigma$ from zero at 40-80\% with $\chi^2/$ndf = 0.24 and 0.06, respectively. Figure~\ref{fig:delv1y} also shows the dependence on $p_\mathrm{T}/n_q$. We reverse the sign of $v_1(y<0)$ before averaging $v_1$ over all $y$ to get the $p_\mathrm{T}/n_q$ dependence. 
For $\Delta q=0$ and $\Delta S=0$, the measured $\Delta v_1$ versus $p_\mathrm{T}/n_q$ fluctuates and the estimated parameters from pol2 ($b_0+b_1x+b_2x^2$) fitting are $b_0 = -0.0085 \pm 0.0093$, $b_1 = 0.0141 \pm 0.0362$ and $b_2 = -0.0002 \pm 0.0334$ with $\chi^2$/ndf =0.246. This yields larger uncertainties in slope. 
Current data thus offer limited constraints when testing the coalescence sum rule in the context of $p_\mathrm{T}/n_q$ dependence.

Fitted $F_\Delta$ slopes for all ($\Delta q,\,\Delta S$) combinations at 10-40\% centrality are listed in Table~\ref{tab:delq_dels} (Appendix~\ref{a1}). For ($\Delta q,\,\Delta S$)=(4/3,~2), our measurement has the best precision, and the increments of $\Delta v_1$ with $|y|$ and $p_T/n_q$ are highly significant. This is evident from the lower panels of Fig.~\ref{fig:delv1y}.
As per the discussion of Fig.~\ref{cartoon}, increased $\Delta v_1$ for $\Delta q \ne 0$ and $\Delta S \ne 0$ is consistent with expectations from EM fields. For 40-80\% centrality, 
our track statistics are lower by a factor of about one-third, and distinguishing between ($\Delta q,\,\Delta S$)$=$(4/3,~2) and (0,~0) becomes more difficult.

In Fig.~\ref{fig:slope_delq_dels}, we display midrapidity $F_\Delta$ for the studied combinations versus $\Delta q$ and $\Delta S$ in 10-40\% Au+Au at $\sqrt{s_{\mathrm{NN}}} = 27$ and 200 GeV. 
We fit the measurements with a first-order polynomial, pol1 (blue dashed lines in Fig.~\ref{fig:slope_delq_dels}), and obtain $10^4 \times F_\Delta = (6\pm28)+(24\pm21)\Delta q$, and $10^4 \times F_\Delta = (17\pm21)+(0\pm15)\Delta q$ for 10-40\% at $\sqrt{s_{\mathrm{NN}}} = 27$ and 200 GeV, respectively. For 40-80\% centrality, $10^4 \times F_\Delta = (130\pm65)+(-77\pm49)\Delta q$, and $10^4 \times F_\Delta = (-16\pm34)+(27\pm25)\Delta q$ at $\sqrt {s_{\mathrm{NN}}} = 27$ and 200 GeV, respectively. At 40-80\% centrality, given the large uncertainties, intercepts and slopes are mostly consistent with zero. Similar conclusions are reached from the $\Delta S$ dependence of $d\Delta v_1/dy$.



In contrast, Fig.~\ref{fig:slope_delq_dels} indicates an overall positive $F_\Delta$. Fitting a constant to the data yields $10^4\times F_\Delta= 39\pm6$ and $17\pm5$ for 10-40\% centrality at 27 and 200 GeV, respectively. We also explored a two-dimensional fit of the measured $F_\Delta$, but no  physics conclusion was possible due to large uncertainties.

If the EM field is responsible for this positive splitting, $F_\Delta$ should be proportional to $\Delta q$. We therefore assume that the observed small deviations from the sum rule are dominated by EM effects,
and perform linear fits with zero intercepts: $F_\Delta=K_{\Delta q}\Delta q$, and $F_\Delta=K_{\Delta S}\Delta S$. Results are shown in Fig.~\ref{fig:slope_delq_dels}, 
 where a $\pm$1$\sigma$ band accompanies each fit line. Fitted slopes $K_{\Delta q}$ and $K_{\Delta S}$ are given in Table~\ref{tab:splitting}. 
The splitting, $F_\Delta$, increases with $\Delta q$ and $\Delta S$. At 10-40\% centrality, the significance of this increase is $4.8\sigma$ for $\Delta q$ and $4.6\sigma$ for $\Delta S$ at 27 GeV. The significance is dominated by the measurements at ($\Delta q$,\,$\Delta S$)$=$(4/3,~2). 
The splitting evidently increases between $\sqrt{s_{\mathrm{NN}}}=$ 200 and 27 GeV, as does the predicted influence of the intense $B$ field~\cite{McLerran:2013hla}. 
Our measurements of positive splitting $d\Delta v_1/dy$ for positive $\Delta q$ is consistent with the expectation of the Hall effect dominating over Coulomb+Faraday in the EM scenario (see Fig.~\ref{cartoon}) as argued in Ref~\cite{Gursoy:2018yai}.

In Fig.~\ref{fig:slope_delq_dels} we plot AMPT~\cite{Nayak:2019vtn} model calculations (string melting mode) and they fail to describe our measurements. Ref.~\cite{Nayak:2019vtn} does not report calculations for $\Omega$ baryons and therefore AMPT in Fig.~\ref{fig:slope_delq_dels} does not cover the full range of the measurements.
AMPT does not include an EM field but incorporates coalescence where only spatial proximity of the coalescing quarks is used and velocity-space proximity is ignored. AMPT predicts negative splitting, opposite to the observed data, indicating a lack of consideration for the crucial physics underlying $v_1$ splitting. One potential factor contributing to this discrepancy could be the EM field. 

Parida and Chatterjee \cite{Parida:2023ldu} reported for the 5 studied combinations,
$\Delta S=-3 \Delta B$ and they suggested the observed splitting might be explained by increasing baryon inhomogeneities in the system even in the absence of EM-field effects. Further studies based on future precision measurements of heavier baryons can probe this interesting effect.

In summary, we report measurements of $v_1(y)$ for multistrange baryons ($\Xi$ and $\Omega$) in Au+Au collisions at $\sqrt{s_{\mathrm{NN}}}=$ 27 and 200 GeV. 
We focus on seven produced particle species: $K^{-}$, $\bar{p}$, $\bar{\Lambda}$, $\phi$, $\overline{\Xi}^{+}$, ${\Omega}^{-}$ and $\overline{\Omega}^{+}$, none of whose constituent quarks is transported from the colliding nuclei, and study the difference (splitting), $F_\Delta=d\Delta v_1/dy$, between pairs of particle combinations with similar quark content but varying electric charge difference $\Delta q$ and strangeness difference $\Delta S$. For $\Delta q = \Delta S =0$, consistency with the coalescence sum rule is observed within errors.
For $\Delta q \ne 0 \ne \Delta S$, a nonzero average $F_\Delta$ is observed with $>5\sigma$ significance. 
This splitting increases with $\Delta q$ (correlated with increasing $\Delta S$). However, this increase is not statistically significant without assuming the coalescence sum rule. Given this assumption, the one-parameter linear fit versus $\Delta q$ and $\Delta S$ yields a positive slope, respectively $4.8\sigma$ and $4.6\sigma$ away from zero for $\sqrt{s_{\mathrm{NN}}}=$ 27 GeV. This splitting is stronger at 27 GeV than at 200 GeV. The AMPT model (where no EM fields are implemented) fails to describe the data. 
These 10-40\% data are therefore consistent with electromagnetic effects, where the positive slope suggests a dominance of Hall effect over Faraday+Coulomb effect. 
A companion analysis~\cite{STAR:2023jdd} concludes that in peripheral collisions, light quark $v_1$ splitting is dominated by Faraday+Coulomb effects, with a flavor dependence between light and strange quarks. Our work, alongside the companion work~\cite{STAR:2023jdd}, including transport quarks, suggests a centrality and flavor-dependent competition between the Hall effect and the Faraday+Coulomb effects.

\section*{Acknowledgment}
We thank the RHIC Operations Group and SDCC at BNL, the NERSC Center at LBNL, and the Open Science Grid consortium for providing resources and support.  This work was supported in part by the Office of Nuclear Physics within the U.S. DOE Office of Science, the U.S. National Science Foundation, National Natural Science Foundation of China, Chinese Academy of Science, the Ministry of Science and Technology of China and the Chinese Ministry of Education, NSTC Taipei, the National Research Foundation of Korea, Czech Science Foundation and Ministry of Education, Youth and Sports of the Czech Republic, Hungarian National Research, Development and Innovation Office, New National Excellency Programme of the Hungarian Ministry of Human Capacities, Department of Atomic Energy and Department of Science and Technology of the Government of India, the National Science Centre and WUT ID-UB of Poland, the Ministry of Science, Education and Sports of the Republic of Croatia, German Bundesministerium f$\ddot{\rm{u}}$r Bildung, Wissenschaft, Forschung and Technologie (BMBF), Helmholtz Association, Ministry of Education, Culture, Sports, Science, and Technology (MEXT), Japan Society for the Promotion of Science (JSPS) and Agencia Nacional de Investigaci\'on y Desarrollo (ANID) of Chile.

\newpage
\appendix
\section{ Supplemental material }
\subsection{Tabulated Results}
\label{a1}

\begin{table*}[th]
\renewcommand{\arraystretch}{1.8}
\begin{tabular}{l l l l l l l l l l l l l}
\hline
\hline
Index & Quark mass && $\Delta q$ && $\Delta S$ & & $\Delta B$ & &  ~~~~~$\Delta v_1$ combination & & ~~$F_{\Delta}\times 10^{4}$ (27 GeV)  & ~$F_{\Delta} \times 10^{4}$ (200 GeV)   \\ 
\hline
1 & $\Delta m=0$ & & $0$ & & $0$ & & $0$ & & ${[\pbar + \ph]}-[\km + \al ]$ & & ~~~~$03\pm43\pm13$ & ~~~$56\pm49\pm41$ \\ 
2 & $\Delta m\approx 0$ & & $1$ & & $2$ & & $-\frac{2}{3}$ & &
${[\al]}-[\frac{1}{3}\om + \frac{2}{3}\pbar]$ & & ~~~~$41\pm25\pm16$ & ~~~$19\pm13\pm01$ \\
3 & $\Delta m\approx 0$ & & $\frac{4}{3}$ & & $2$ & & $-\frac{2}{3}$ & &
${[\al]}-[\km + \frac{1}{3}\pbar]$ & & ~~~~$39\pm07\pm03$ & ~~~$16\pm05\pm03$\\
4 & $\Delta m= 0$ & & $2$ & & $6$ & & $-2$ & &
${[\op]}-[\om]$ & & ~~~~$83\pm130\pm25$ & ~~~$35\pm58\pm54$ \\
5 & $\Delta m\approx 0$ & & $\frac{7}{3}$ & & $4$ & & $-\frac{4}{3}$ & &
${[\ks]}-[\km + \frac{1}{3}\om]$ & & ~~~~$64\pm36\pm19$ & ~~~$26\pm20\pm04$\\
\hline
\hline
\end{tabular}
\caption{Linearly-independent combinations of hadrons whose constituent quarks are all produced. For all combinations, $\Delta m$ for constituent quarks is zero or near zero, while the charge ($\Delta q$), strangeness ($\Delta S$) and baryon number ($\Delta B$) differences vary as tabulated. Here $\Delta S$ and $\Delta B$ are related by $\Delta S=-3 \Delta B$. The measured $F_{\Delta} = d\Delta v_1/dy$ for the 10-40\% centrality bin is also shown. The listed errors are absolute statistical and systematic uncertainties, in that order. }
\label{tab:delq_dels}
\end{table*}

\begin{table}[th]
\renewcommand{\arraystretch}{1.5}
\begin{tabular}{l l l l l l l}
\hline
\hline
Parameters & $\sqrt {s_{\mathrm{NN}}} = 27$ GeV & $\sqrt {s_{\mathrm{NN}}} = 200$ GeV & Centrality  \\ 
\hline
$K_{\Delta q}(\times10^{4})$ & ~$29.0\pm4.2\pm3.7$ & ~$12.0\pm3.3\pm2.6$ & ~~ 10-40\% \\ 
$K_{\Delta S}(\times10^{4})$ & ~$19.0\pm2.8\pm2.5$ & ~$7.5\pm2.1\pm1.4$ & ~~ 10-40\%\\ 
$K_{\Delta q}(\times10^{4})$ & ~$19.0\pm15.0\pm9.8$ & ~$15.0\pm5.5\pm3.0$ & ~~ 40-80\%\\
$K_{\Delta S}(\times10^{4})$ & ~$13.0\pm7.5\pm6.5$ & ~$9.5\pm3.6\pm1.9$ & ~~ 40-80\% \\
\hline
\hline
\end{tabular}
\caption{Fit parameters for $F_\Delta=K_{\Delta q}\Delta q$, and $F_\Delta=K_{\Delta S}\Delta S$ at 10-40\% and 40-80\% centrality. }
\label{tab:splitting}
\end{table}

\subsection{Results for 40-80\% centrality}

\label{a2}

\begin{figure}[htb]
    \centering
   \includegraphics[width=0.49\textwidth]{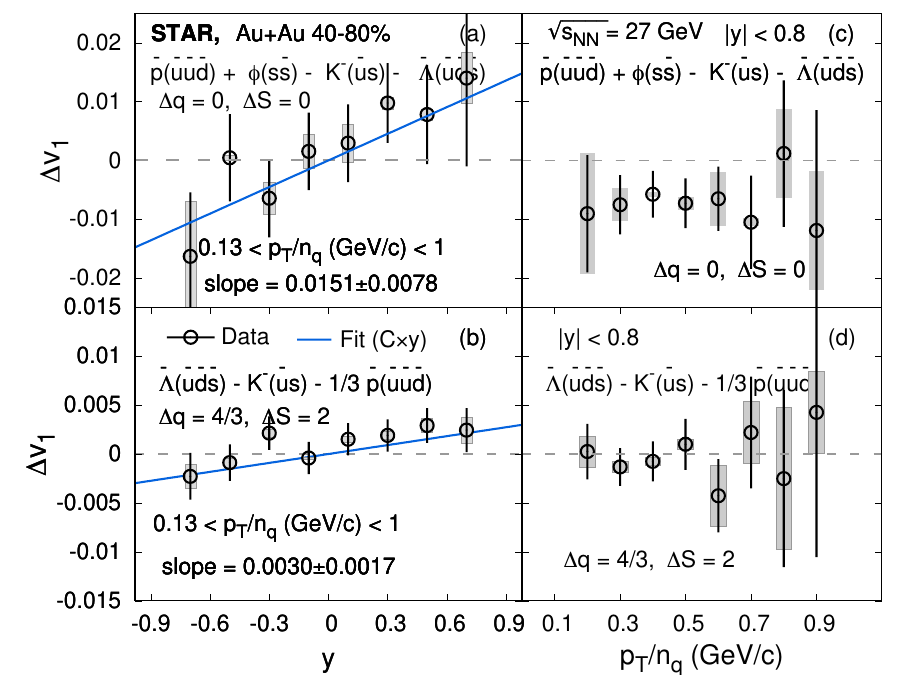}
   \caption{The $\Delta v_1$ as a function of rapidity (a, b) and $p_\mathrm{T}/n_q$ (c, d) for ($\Delta q,\,\Delta S$)=(0,~0) and (4/3,~2) in 40-80\% Au+Au at $\sqrt{s_{\mathrm{NN}}}=27$ GeV. The vertical bars and shaded bands denote statistical and systematic uncertainties, respectively.}
   \label{fig:delv1y-peri}
\end{figure}

In Fig.~\ref{fig:delv1y-peri}, we present the measured $\Delta v_1(y)$ and the corresponding linear fits for $\Delta q=0$ and $\Delta S=0$ and for $\Delta q=4/3$ and $\Delta S=2$ in 40-80\% centrality in Au+Au collisions at $\sqrt {s_{\mathrm{NN}}} = 27$ GeV. 

\begin{figure}[htb]
    \centering
   \includegraphics[width=0.49\textwidth]{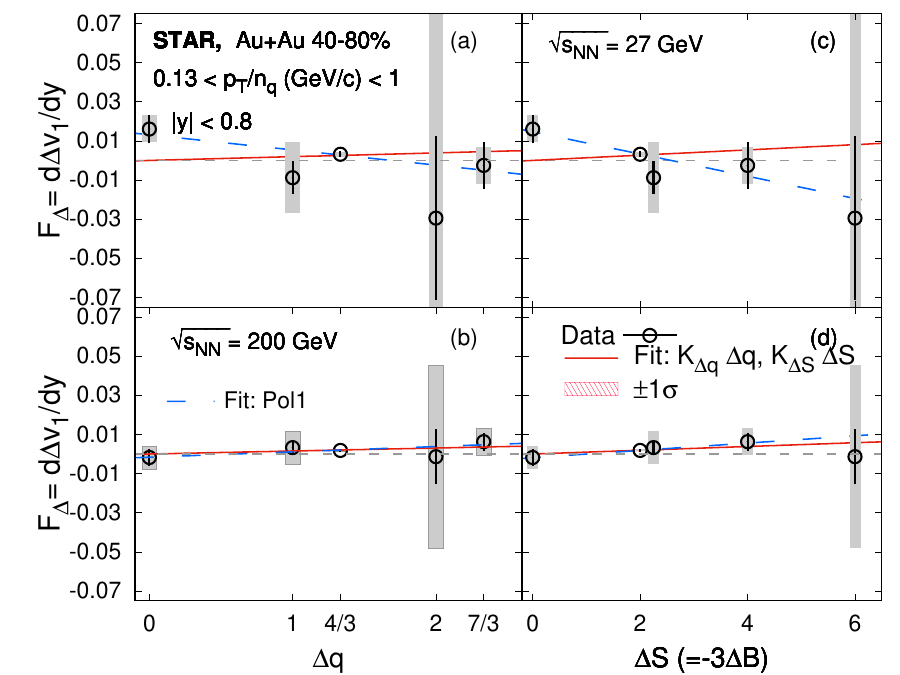}
   \caption{Midrapidity $\Delta v_1$ slope versus $\Delta q$ (a, b), $\Delta S$ (c, d) in 40-80\% Au+Au at 27 (a, c) and 200 (b, d) GeV, respectively. The dashed curves show AMPT  calculations. 
    The vertical bars and shaded bands denote statistical and systematic uncertainties, respectively. There are two degenerate points at $\Delta S=2$; one is displaced horizontally for better visualization.}
   \label{fig:slope-peri}
\end{figure}

Figure~\ref{fig:slope-peri} displays $F_\Delta$ at midrapidity for the studied combinations as a function of $\Delta q$ and $\Delta S$ in 40-80\% central Au+Au collisions at $\sqrt {s_{\mathrm{NN}}} = 27$ and 200 GeV. The measurements are dominated by the uncertainties arising from the low abundance of massive hadrons used in the analysis.

\bibliography{main}

\clearpage
\end{document}